\documentclass[conference]{IEEEtran}
\IEEEoverridecommandlockouts
\usepackage{cite}
\usepackage{amsmath,amssymb,amsfonts}
\usepackage{algorithmic}
\usepackage{graphicx}
\usepackage{textcomp}
\usepackage{xcolor}
\usepackage[linesnumbered,ruled]{algorithm2e}
\usepackage{float}                  %
\usepackage{subfig}                 %
\usepackage{overpic}                %
\usepackage{caption}
\usepackage{geometry}
\def\BibTeX{{\rm B\kern-.05em{\sc i\kern-.025em b}\kern-.08em
    T\kern-.1667em\lower.7ex\hbox{E}\kern-.125emX}}
\begin{document}

\title{Specific Emitter Identification Based on Joint Variational Mode Decomposition\\
}

\author{\IEEEauthorblockN{Xiaofang Chen\textsuperscript{1}, Wenbo Xu\textsuperscript{1}, Yue Wang\textsuperscript{2}, Yan Huang\textsuperscript{3}}
\IEEEauthorblockA{1.\textit{Key Laboratory of Universal Wireless Communication, Beijing University of Posts and Telecommunications}, Beijing, China\\
2. \textit{Department of Computer Science, Georgia State University}, Atlanta, GA\\
3. \textit{Department of Software Engineering and Game Development, Kennesaw State University}, Marietta, GA 30060, USA\\
E-mails: \{chenxf, xuwb\}@bupt.edu.cn, ywang182@gsu.edu, yhuang24@kennesaw.edu}
}

\maketitle

\begin{abstract}
Specific emitter identification (SEI) technology is significant in device administration scenarios, such as self-organized networking and spectrum management, owing to its high security. 
For nonlinear and non-stationary electromagnetic signals, SEI often employs variational modal decomposition (VMD) to decompose the signal in order to effectively characterize the distinct device fingerprint. 
However, the trade-off of VMD between the robustness to noise and the ability to preserve signal information has not been investigated in the current literature.
Moreover, the existing VMD algorithm does not utilize the stability of the intrinsic distortion of emitters within a certain
temporal span, consequently constraining its practical applicability in SEI.
In this paper, we propose a joint variational modal decomposition (JVMD) algorithm, which is an improved version of VMD by simultaneously implementing modal decomposition on multi-frame signals. 
The consistency of multi-frame signals in terms of the central frequencies and the inherent modal functions (IMFs) is exploited, which effectively highlights the distinctive characteristics among emitters and reduces noise.
Additionally, the complexity of JVMD is analyzed, which is proven to be more computational-friendly than VMD.
Simulations of both modal decomposition and SEI that involve real-world datasets are presented to illustrate that when compared with VMD, the JVMD algorithm improves the accuracy of device classification and the robustness towards noise. 
\end{abstract}


\section{Introduction}
Specific emitter identification (SEI) is a technique used for verifying the safety of emitters at the physical layer \cite{qian2021specific}. It distinguishes emitters by extracting radio frequency fingerprints (RFFs) that result from inherent defects in the emitter. SEI is widely applied in fields like military countermeasures and cognitive radio. 
Additionally, SEI enjoys lower costs and resources when compared with traditional cryptographic security management techniques. With the exponential growth of the number of wireless devices, such advantages have invoked the interest of scholars to study its potential application in the Internet of Things (IoT) and smart living \cite{xu2022lightweight}.

SEI generally consists of three steps: preprocessing, feature extraction, and classification. Among these steps, feature extraction is the most crucial one in current SEI research. 
Three types of features are usually studied, including signal parameter statistical features \cite{hua2018accurate}, mechanism model features \cite{polak2015wireless}, and signal transform domain statistical features \cite{zhang2016specific}, among which the last one is mostly utilized for either nonlinear or non-stationary signals in order to effectively characterize the signal.
To extract features in the transform domain, signal decomposition algorithms in the time-frequency domain have emerged with three prominent algorithms, i.e., empirical modal decomposition (EMD) \cite{rehman2010multivariate}, intrinsic time scale decomposition (ITD) \cite{lang2019direct}, variational modal decomposition (VMD) \cite{dragomiretskiy2013variational, WOS:000399828500007, 8890883}.
VMD, proposed by Dragomiretskiy \it{et al}., \rm{not} only addresses the modal aliasing issue that appeared in EMD but also enjoys the advantage of low complexity, prompting it to be a widely adopted technique in current SEI studies \cite{dragomiretskiy2013variational}.


In recent years, VMD has been introduced into SEI due to its advantages. 
The authors in \cite{satija2018specific} realize SEI with the VMD technique and show favorable recognition accuracy, where the spectral attributes of each electromagnetic signal component are studied. 
Su \it{et al.} \rm{employ} VMD to extract temporal and frequency domain features of the signal to enhance the distinction among transmitters \cite{su2023specific}.
On the other hand, \cite{gok2020new} demonstrates the potential of VMD in radar transmitter identification by using it as a feature extraction technique to decompose the envelope and instantaneous frequency of radar pulses into components with compact support in the frequency domain.
To further improve the classification accuracy of SEI, various approaches have emerged that combine VMD with advanced technologies such as deep learning \cite{rong2023rffsnet, he2020cooperative}. 

However, the current literature on VMD-based SEI generally ignores the noise as well as the SEI-specific conditions and characteristics.
Firstly, in terms of noise, existing VMD algorithms and their improved versions tend to reduce noise by adjusting their penalty factors \cite{dragomiretskiy2013variational, WOS:000399828500007, 8890883}. 
Nevertheless, a larger penalty factor inevitably leads to the loss of signal details, thus impeding the utilization of subtle intrinsic distortions in SEI.
Secondly, those slight intrinsic distortions remain unchanging from frame to frame within a certain temporal span \cite{9353800, 9229093}, which is not considered in current VMD-based SEI research.
It will be positive to take into account both the noise immunity and stability of the slight intrinsic distortion of the emitter.

To address the aforementioned problems, this paper proposes the joint variational modal decomposition (JVMD) algorithm for SEI, which enjoys enhanced noise robustness and superior accuracy compared to the conventional VMD algorithm. Additionally, a comparative analysis and simulational results are presented to confirm the low complexity and effectiveness of JVMD. The main contributions of this paper are summarized as follows:
\begin{itemize}
\item A SEI scheme based on JVMD is proposed, where JVMD first extracts some significant features, and then the dictionary learning technique is employed to realize emitter identification. During feature extraction, we incorporate the channel noise into the system model rather than ignoring it and explore the property that the inherent modal functions (IMFs) and center frequencies remain constant among consecutive frames. Both of these two strategies contribute to highlighting the RFFs in the signal, which enables the superiority of JVMD when compared with the traditional VMD algorithm.
\item The complexity of JVMD is analyzed. The result demonstrates that the complexity of JVMD is lower than that of VMD. Moreover, the complexity advantage becomes more significant as the number of joint signal frames increases.
\item Simulations demonstrate the superiority of the JVMD algorithm.  Through tone separation simulations, we first illustrate the robustness of the JVMD algorithm to noise. 
Then the simulations of the overall SEI scheme verify the ability of the JVMD algorithm to enhance the accuracy of transmitter classification.
\end{itemize}

The rest of this paper is organized as follows: Section II briefly presents the SEI system model and VMD algorithm. The proposed JVMD-based SEI scheme and a comprehensive analysis of the complexity of both VMD and JVMD algorithms are described in Section III. Section IV discusses the comparative experimental results on a real-world dataset. Section V summarizes this paper.
\section{Background}
\begin{figure}
 \centerline{\includegraphics[width=3.5in]{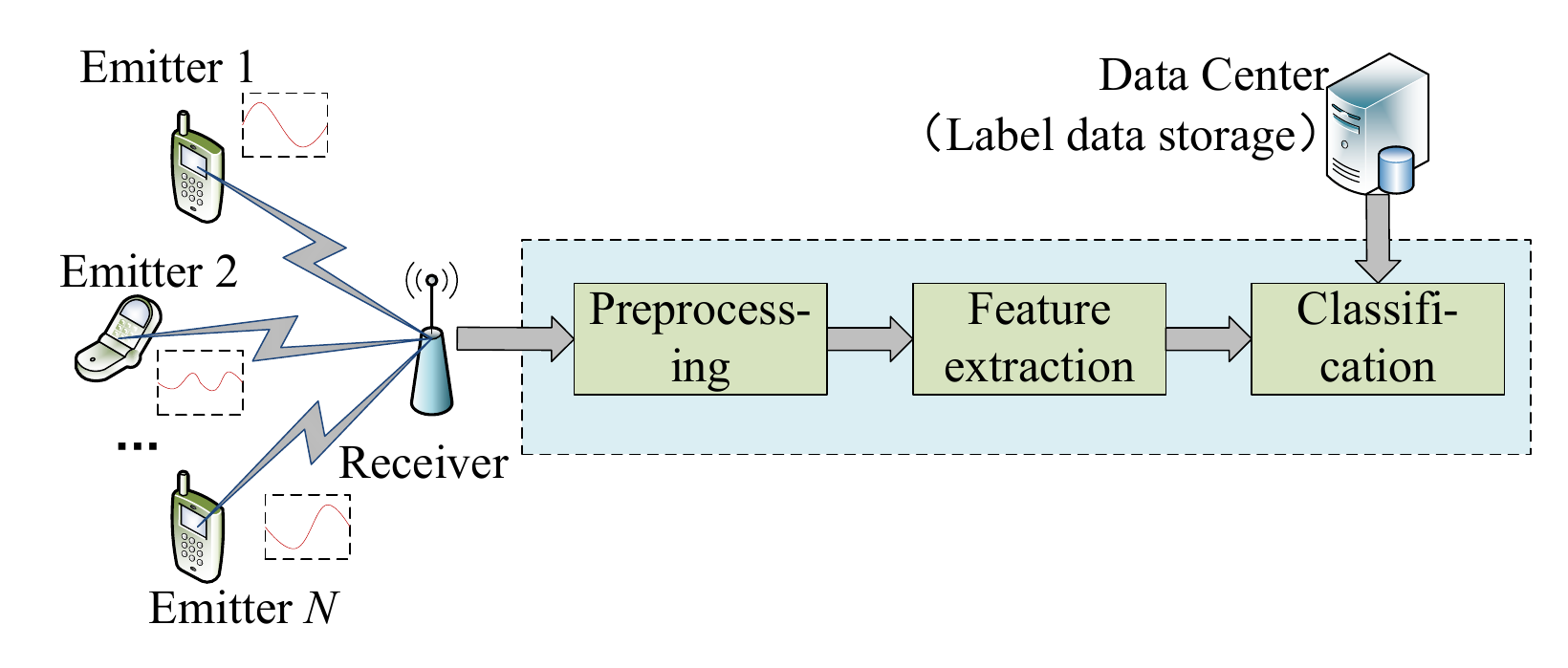}}
\captionsetup{font={footnotesize}}
  \caption{The SEI system model.}
  \label{system}
\end{figure}
In this section, we first present the SEI system model, and then briefly introduce the traditional VMD algorithm.
\subsection{SEI System model}
Fig. \ref{system} gives the system model of a typical SEI scenario.
Assume that there are $N$ emitters, one receiver, and a data center that stores the labeled signals of these emitters in the SEI system.
As shown in Fig. \ref{system}, the receiver acquires the electromagnetic signal, where the signals from different emitters carry the unique and distinguishable intrinsic distortion information of their own emitter.
To facilitate the subsequent processing, the receiver first undertakes some signal preprocessing, such as normalization, domain transformation, etc. 
After that, feature extraction is implemented with a predefined method. 
Finally, the extracted features are classified with the aid of the training labeled data in the data center.

\subsection{Variational Mode Decomposition}
VMD obtains several sub-signals, i.e., IMFs, where each IMF is estimated by solving the frequency domain variational optimization problem. To begin with, the VMD assumes that the IMFs are AM-FM signals, and the $k$th IMF $s_k(t)$ can be expressed as
\begin{equation}
s_k(t)=a_k(t)\cos(\phi_k(t)),
\end{equation}
where the signal envelope $a_k(t)$ is non-negative, and the phase $\phi_k(t)$ is non-decreasing.
Furthermore, VMD introduces analytic signals to maintain only the positive frequency components without loss of information. Assume $s_k(t)$ is a real signal, then its analytic signal is written as
\begin{equation}
z_k(t)=s_k(t)+j\mathcal{H}(s_k(t)).
\end{equation}
In the above equation, $\mathcal{H(\cdot)}$ denotes the Hilbert transform and its system transfer function $h(t)$ is
\begin{equation}
h(t)=\delta(t)+\frac{j}{\pi t}.
\end{equation}
Finally, VMD assumes that all sub-signals are narrow-band that concentrated on their center frequencies, and it formulates the following constrained optimization problem \cite{dragomiretskiy2013variational},
\begin{equation}
\begin{aligned}
\min_{{s_k},{\omega_k}} \quad & \sum_{k=1}^K\left\|\partial_t{\left[\left(\delta(t)+\frac{j}{\pi t}\right)*s_k(t)\right]e^{-j\omega_kt}}\right\|_2^2 & \\
\mbox{s.t.}\quad
&\sum_{k=1}^Ks_k(t)=y(t),
\end{aligned}\label{op}
\end{equation}
where $K$ is the total number of signal components, $\omega_k$ is the center frequencies of the $k$th component, and $y(t)$ is the received signal.

By converting the constrained optimization problem in (\ref{op}) into an unconstrained optimization problem, the augmented Lagrangian function is obtained as follows,
\begin{equation}
\begin{aligned}
\mathcal{L}\left({s_k(t)},{\omega_k},\lambda(t)\right):=
\left\|y(t)-\sum_{i=1}^Ks_i(t)+\frac{\lambda(t)}{2}\right\|_2^2-\left\|\frac{\lambda^2(t)}{4}\right\|_2^2\\
+\alpha\sum_{k=1}^K\left\|\partial_t{\left[\left(\delta(t)+\frac{j}{\pi t}\right)*s_k(t)\right]e^{-j\omega_kt}}\right\|_2^2,
\end{aligned}\label{eq5}
\end{equation}
where the $\alpha$ is the penalty factor and $\lambda(t)$ is the Lagrangian multiplier.
The saddle point of (\ref{eq5}) can be easily obtained by the Alternating Direction Multiplier Method (ADMM), where the $(n+1)$th iteration is implemented as follows,
\begin{figure}[htbp]
  \centerline{\includegraphics[width=3.5in]{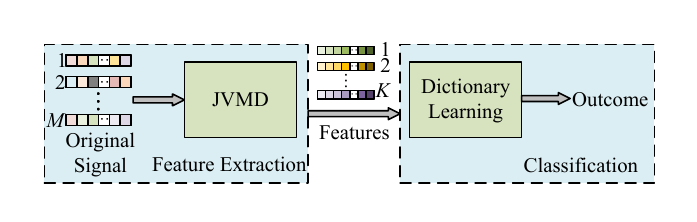}}
  \captionsetup{font={footnotesize}}
  \caption{The overall SEI scheme based on JVMD.}
  \label{scheme}
\end{figure}
\begin{itemize}
\item The updating equation for $s_k(t)$: 
\begin{equation}\label{eq6}
S_k^{(n+1)}(\omega)=\frac{Y(\omega)-\sum_{i\neq k}^{K}S_i(\omega)+\frac{\hat{\lambda}^{(n)}(\omega)}{2}}{1+2\alpha(\omega-\omega_k)^2}.
\end{equation}
\item The updating equation for $\omega_k$:
\begin{equation}\label{w_updata}
\omega_k^{(n+1)}=\frac{\int_0^\infty\omega|S_k(\omega)|^2\,d\omega}{\int_0^\infty|S_k(\omega)|^2\,d\omega}.
\end{equation}
\item The updating equation for $\lambda_k(t)$:
\begin{equation}\label{eq8}
\hat{\lambda}^{(n+1)}(\omega)=\hat{\lambda}^{(n)}(\omega)+\tau(Y(\omega)-\sum_{i=1}^{K}S^{(n+1)}_i(\omega),
\end{equation}
where $\tau$ is the update parameter of $\lambda(t)$.
\end{itemize}
The $S_k(\omega)$, $Y(\omega)$, and $\hat{\lambda}(\omega)$ in the above updating equations are the transformation version of $s_k(t)$, $y(t)$, and $\lambda(t)$ in the frequency domain, respectively. 
\section{Proposed JVMD-based SEI Scheme}
In this section, we propose the JVMD-based SEI scheme and provide the complexity analysis of VMD and JVMD. We first provide an overview of the overall scheme, followed by a detailed description of the feature extraction with JVMD, and classification steps in the scheme. Finally, we provide the complexity analysis of both the JVMD and VMD algorithms.
\subsection{Overall Scheme of SEI}
Fig. \ref{scheme} demonstrates the overall JVMD-based SEI scheme. As shown in this figure, the $M$-frame signal is first decomposed into $K$ sub-signals (i.e., features) jointly by JVMD to complete the feature extraction step. Then, the sub-signals and labeled fingerprints stored in the data center are used in the dictionary learning module to perform the classification, thus obtaining the label of these $M$-frame signals.

\subsection{Feature extraction--JVMD}
The basis of accurate SEI is to extract the distinct features, i.e., the inherent distortion, of each transmitter. The traditional VMD only relies on the information from one signal frame and enhances its robustness to external noise by increasing $\alpha$ in (\ref{eq5}), which unavoidably loses more detailed information about the signal. Considering the fact that the inherent distortion of the transmitter remains almost constant among several consecutive frames \cite{9353800, 9229093}, the joint exploitation of these frames inevitably highlights the distortion and consequently contributes to SEI. To obtain such benefits, we propose the JVMD algorithm with $M$ signal frames by extending the problem in (\ref{op}) to the following form
\begin{equation}
\begin{aligned}
\min_{{s_k},{\omega_k}} \quad & \sum_{k=1}^K\left\|\partial_t{\left[\left(\delta(t)+\frac{j}{\pi t}\right)*s_k(t)\right]e^{-j\omega_kt}}\right\|_2^2 +\varepsilon\sum_{j=1}^M\left\|b_j(t)\right\|^2_2 & \\
\mbox{s.t.}\quad
&\sum_{k=1}^Ks_k(t)+b_j(t)=y_j(t), j=1,2,\cdots, M,
\end{aligned}\label{eq9}
\end{equation}
where $b_j(t)$ is the noise of the $j$th signal frame, $\varepsilon$ is a weighting factor, and the definitions of the remaining variables are the same as those in Section II. 

By utilizing the augmented Lagrangian function, we transform the above equation-constrained optimization problem to an unconstrained one shown in (\ref{LP}).
\begin{figure*}
\begin{equation}
\begin{aligned}
\mathcal{L}\left({s_k(t)},{\omega_k},b_j(t),\lambda_j(t) \right):=\alpha\left\{\sum_{k=1}^K\left\|\partial_t{\left[\left(\delta(t)+\frac{j}{\pi t}\right)*s_k(t)\right]e^{-j\omega_kt}}\right\|_2^2 +\varepsilon\sum_{j=1}^M\left\|b_j(t)\right\|^2_2\right\} &\\
-\sum_{j=1}^M\left\|\frac{\lambda_j^2(t)}{4}\right\|_2^2
+\sum_{j=1}^M\left\|y_j(t)-\sum_{i=1}^Ks_i(t)-b_j(t)+\frac{\lambda_j(t)}{2}\right\|_2^2.
\end{aligned}\label{LP}
\end{equation}
 \hrulefill
\end{figure*}
To deal with the problem in (\ref{LP}), ADMM is used in an iterative manner to solve four subproblems. The updated details are given as follows:
\begin{itemize}
\item Minimization with respect to $s_k(t)$:\\
In order to update $s_k(t)$, we first write the optimization subproblem associated with $s_k(t)$ according to (\ref{LP}):
\begin{equation}
\begin{aligned}
s_k^{(n+1)}=\mathop{\arg\min}\limits_{s_k(t)}
\alpha\left\|\partial_t\left[\left(\delta(t)+\frac{j}{\pi t}\right)*s_k(t)\right]e^{-j\omega_kt}\right\|^2_2\\
+\sum_{j=1}^M\left\|y_j(t)-\sum_{k=1}^Ks_k(t)-b_j(t)+\frac{\lambda_j(t)}{2}\right\|^2_2.
\end{aligned}\label{eq11}
\end{equation}
The optimization in (\ref{eq11}) is complicated due to the convolution operation. To make the problem solvable, we convert it to the frequency domain and get
\begin{equation}
\begin{aligned}
S_k^{(n+1)}=\mathop{\arg\min}\limits_{S_k(\omega)}{\alpha\left\|j(\omega-\omega_k)\left(1+\rm{sgn}(\omega)\right)S_k(\omega)\right\|^2_2} &\\
+\sum_{j=1}^M\left\|Y_j(\omega)-\sum_{i=1}^KS_i(\omega)-\hat{b}_j(\omega)+\frac{\hat{\lambda}_j(\omega)}{2}\right\|^2_2.
\end{aligned}\label{eq12}
\end{equation}
Note that the norm sign and the $\rm{sgn}(\cdot)$ function in (\ref{eq12}) disable the optimization due to their Non-differentiability.
To overcome such difficulty, we rewrite the above equation in the form of a positive frequency integral as follows,
\begin{equation}
\begin{aligned}
S_k^{(n+1)}=\mathop{\arg\min}\limits_{S_k(\omega)}{\int_0^\infty4\alpha(\omega-\omega_k)^2\left|S_k(\omega)\right|^2} \\
+2\sum_{j=1}^M\left|Y_j(\omega)-\sum_{i=1}^KS_i(\omega)-\hat{b}_j(\omega)+\frac{\hat{\lambda}_j(\omega)}{2}\right|^2\,d\omega.
\end{aligned}\label{eq13}
\end{equation}
Finally, by calculating the derivation of (\ref{eq13}), $S_k(\omega)$ is updated as
\begin{equation}
\begin{aligned}
S_k^{(n+1)}(\omega)=\frac{\sum_{j=1}^M\left(U_j(\omega)-\sum_{i\neq k}^KS_i(\omega)\right)}{M+2\alpha(\omega-\omega_k)^2}
\end{aligned},\label{s_updata}
\end{equation}
where 
\begin{equation}
U_j(\omega)=Y_j(\omega)-\hat{b}_j(\omega)+\frac{\hat{\lambda}_j(\omega)}{2}.
\end{equation}

\item Minimization with respect to $\omega_k$:\\
The subproblem term associated with $\omega_k$ in (\ref{LP}) is the same as in the traditional VMD algorithm, hence the update formula for $\omega_k$ in JVMD remains the same as (\ref{w_updata}) in Section II.

\item Minimization with respect to $b_j(t)$:\\
The update of $b_j(t)$ is similar to that of $s_k(t)$. We first write the optimization subproblem associated with $b_j(t)$ according to (\ref{LP}), and then convert it to the frequency domain. The optimization subproblem in the form of a positive frequency integral is obtained as
\begin{equation}
\begin{aligned}
\hat{b}_j^{(n+1)}=\mathop{\arg\min}\limits_{\hat{b}_j(\omega)}{\int_0^{\infty}}2\alpha\varepsilon\left|\hat{b}_j(\omega)\right|^2+\\
2\left|Y_j(\omega)-\sum_{i=1}^KS_i(\omega)-\hat{b}_j(\omega)+\frac{\hat{\lambda}_j(\omega)}{2}\right|^2\,d\omega.
\end{aligned}
\end{equation}
Similarly, this optimization problem can be solved by taking partial derivatives, and then the updated $\hat{b}_j(\omega)$ is obtained as
\begin{equation}
\begin{aligned}
\hat{b}_j^{(n+1)}(\omega)=\frac{Y_j(\omega)-\sum_{i=1}^KS_i(\omega)+\frac{\hat{\lambda}_j(\omega)}{2}}{1+\alpha\varepsilon}
\end{aligned}.\label{b_updata}
\end{equation}

\item Minimization with respect to $\lambda(t)$:\\
Following the updating strategy for $s_k(t)$, the optimization subproblem for $\hat{\lambda}_j(\omega)$ is obtained as
\begin{equation}
\begin{aligned}
\hat{\lambda}_j^{(n+1)}=\mathop{\arg\min}\limits_{\hat{\lambda}_j(\omega)}\int_0^\infty-2\left|\frac{\hat{\lambda}^2_j(\omega)}{4}\right|^2+\\
2\left|Y_j(\omega)-\sum_{i=1}^KS_i(\omega)-\hat{b}_j(\omega)+\frac{\hat{\lambda}_j(\omega)}{2}\right|^2\,d\omega.
\end{aligned}
\end{equation}
Then, it is easy to get the updated frequency version of $\lambda(t)$ :
\begin{equation}
\begin{aligned}
\hat{\lambda}_j^{(n+1)}(\omega)=\sqrt[3]{4\left(Y_j(\omega)-\sum_{i=1}^KS_i(\omega)-\hat{b}_j(\omega)+\frac{\hat{\lambda}_j^{(n)}(\omega)}{2}\right)}
\end{aligned}.\label{lambda_updata}
\end{equation}
\end{itemize}

The procedures of the JVMD algorithm are summarized in Algorithm 1.
The iterative updating consists of two main parts, where one is to update variables $S_k(t)$ and $\omega_k$, and the other one is to update auxiliary variables $\hat{b}_j(t)$ and $\hat{\lambda}(t)$. 

\begin{algorithm}
\DontPrintSemicolon
  \SetAlgoLined
  \KwIn {The received signal $y_j(t), j=1,2,\cdots, M$; The parameters of JVMD $\alpha$, $\varepsilon$; The maximum number of iterations $N$; The convergence threshold $\xi$. }
  \KwOut {$S_k(\omega)$, $\omega_k$, $k=1,\cdots, K$.}
  initialization\;
  $\{s_k^{(1)}(t)\}, \{\omega_k^{(1)}\}, \{b_j^{(1)}(t)\}, \lambda_j^{(1)}(t)\gets 0$; $n \gets 1$; $\kappa \gets \infty$;\;
  \While{$n<N$ and $\kappa>\xi$}{
   $n\gets n+1$ \;
   \For{k=1:K}{
   Update $S_k^{(n)}(\omega)$ by (\ref{s_updata}); \;
   Update $\omega_k^{(n)}$ by (\ref{w_updata}); \;
   }
   \For{j=1:M}{
   Update $\hat{b}_j^{(n)}(\omega)$ by (\ref{b_updata}); \;
   Update $\hat{\lambda}_j^{(n)}(\omega)$ by (\ref{lambda_updata}); \;
   }
   $\kappa=\sum_{i=1}^K\left\|S_k^{(n)}(\omega)-S_k^{(n-1)}\right\|^2_2/\left\|S_k^{(n)}(\omega)\right\|^2_2$; 
  }
  \caption{The procedures of the JVMD}
\end{algorithm}
\subsection{Classification--Dictionary Learning}
In this paper, considering that SRC \cite{qian2021specific} and Lable Consistent K-SVD (LC-KSVD) \cite{tang2020dictionary} are two typical dictionary learning algorithms, we use them to realize classification. 
Specifically, SRC is a classification algorithm based on sparse theory,  where the training signals form the dictionary, and the signal to be classified acts as the observation signal. By employing various sparse reconstruction methods, SRC determines the category of the signal based on the derived sparse coefficients.
LC-KSVD, on the other hand, considers the label consistency and formulates an appropriate dictionary.
By doing so, LC-KSVD induces greater correlation in the sparse codes of the same class of signals and more pronounced differences in the sparse codes of different classes of signals.
\begin{figure*}[htbp]
	\centerline{
	\subfloat[SNR=30dB]{\includegraphics[width=.32\linewidth]{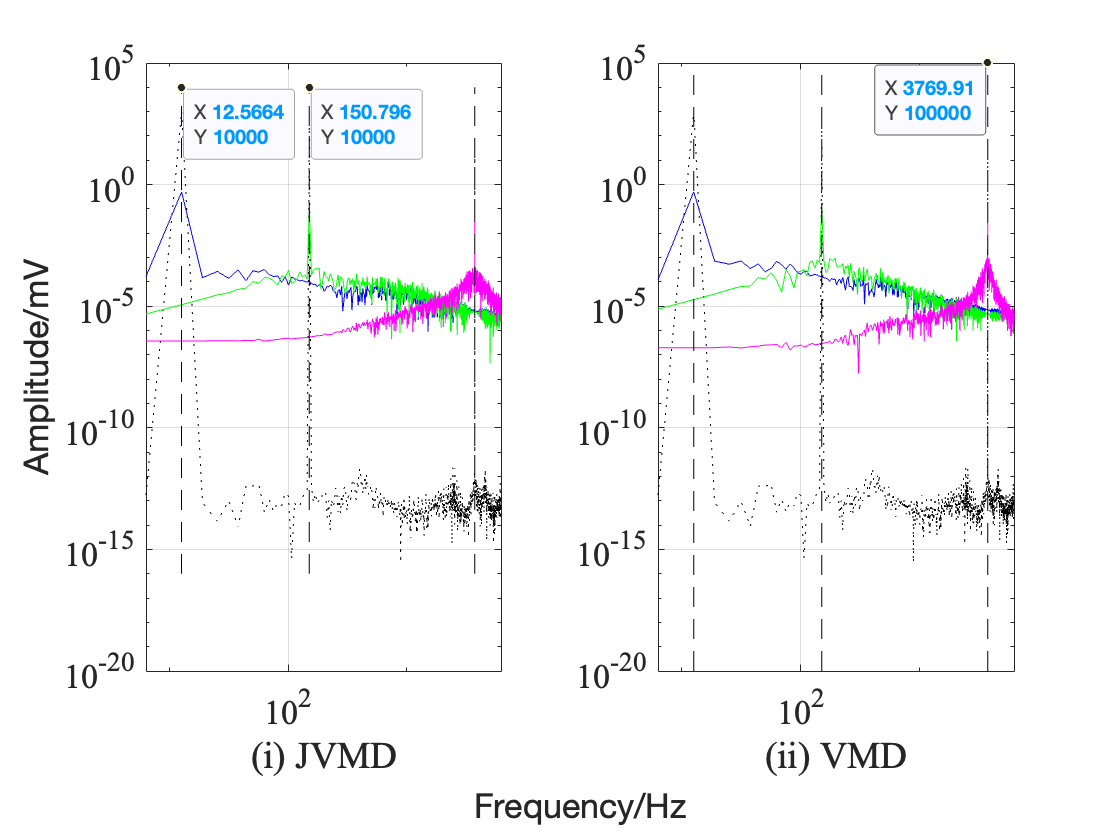}}\hspace{5pt}
	\subfloat[SNR=10dB]{\includegraphics[width=.32\linewidth]{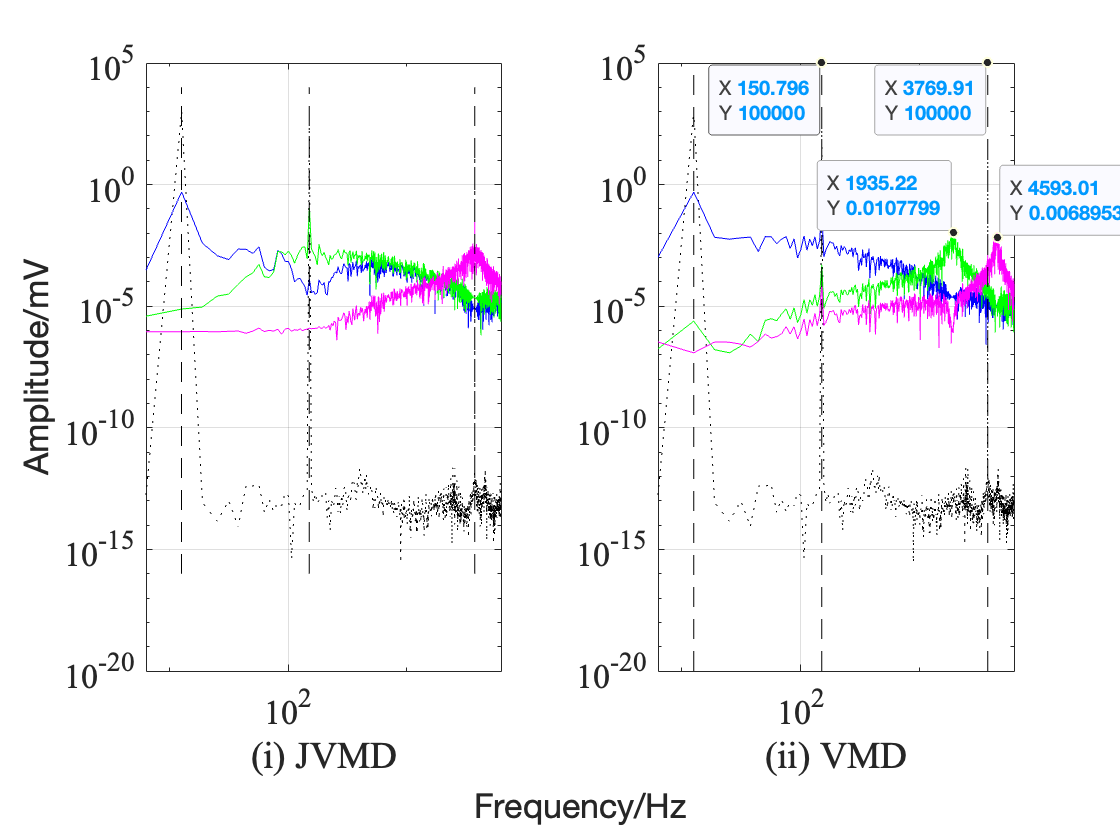}}
	\subfloat[The center frequency relative error]{\includegraphics[width=.32\linewidth]{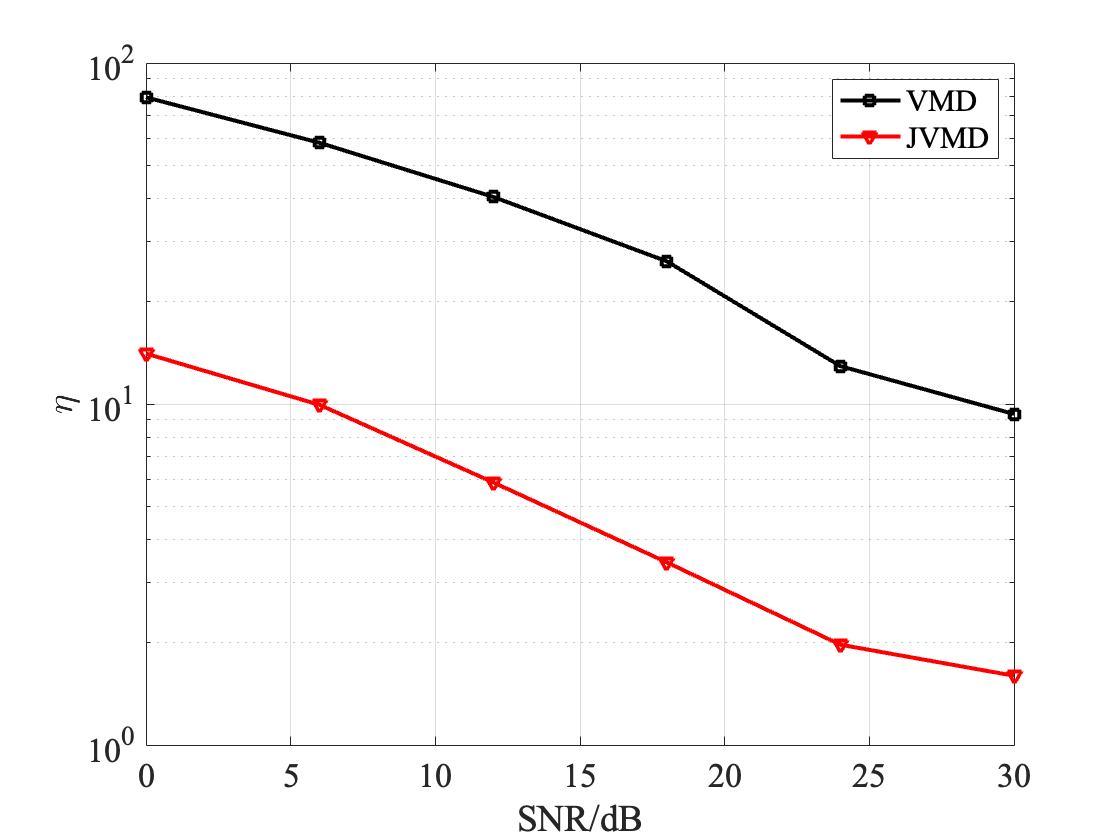}}}
	\captionsetup{font={footnotesize}}
	\caption{Tone separation simulation with $\alpha=2000$, $L=1500$, $M=8$, $\xi=10^{-7}$, where (a) and (b) plot the power spectral density of each tone with different SNRs, (c) exhibits the center frequency relative error of VMD and JVMD with varying SNRs.}\label{TS}
\end{figure*}
\subsection{Complexity Analytics}
This subsection discusses the complexity of VMD and JVMD. For the VMD algorithm, the decomposition of each frame is performed independently. The computational effort is mainly concentrated on the update iterations of (\ref{eq6}) to (\ref{eq8}), which has the complexity of $O(K^2L)$. Consequently, if performing VMD on $M$ frames, the total complexity is $O(MK^2L)$.
On the other hand, JVMD decomposes $M$ frames jointly. As demonstrated in Algorithm 1, the main complexity lies in the two inner loop parts, i.e., Steps 5 to 8 and Steps 9 to 12, whose complexities are $O(K^2L)$ and $O(MKL)$, respectively. Therefore, the complexity of JVMD is the sum of the complexity of these two inner loops, i.e., $O(K^2L+MKL)$.

Note that when $M=1$, the complexity of VMD and JVMD are at the same level. However, when $M>1$, the complexity of JVMD is lower than that of VMD, and this complexity advantage becomes more obvious as $M$ increases.
\section{Simulation Results}
In this section, some simulations are provided to show the performance of the proposed SEI scheme. 
To verify the robustness of JVMD, we first give the tone separation simulations with a mixed signal consisting of three harmonics. Then, the proposed JVMD-based SEI scheme is implemented on a real-world dataset to verify its efficiency.


\subsection{JVMD for Tone Separation}
In the tone separation simulation, the input signal $y(t)$ consists of a mixture of three signals with different tones, such that
\begin{equation}
y(t)=\cos(4\pi t)+\frac{1}{4}\cos(48\pi t)+\frac{1}{16}\cos(1200\pi t)+w(t),
\end{equation}
where $w(t)$ is the additive Gaussian white noise (AWGN).
The main objective of the simulation is to accurately separate the three tone components in the above equation. Therefore, we measure the accuracy of tone separation by the center frequency relative error $\eta$, which is defined as follows,
\begin{equation}
\eta=\left\|\frac{\bf{f}-\bf{f}_c}{\bf{f}_c}\right\|^2_2,
\end{equation}
where $\bf{f}$ and $\bf{f}_c$ are the estimated and actual center frequency, respectively.

Fig. \ref{TS} illustrates the results of tone separation obtained by the JVMD and VMD algorithms at various signal-to-noise ratios (SNRs). In this figure, the parameters of both algorithms in Section II and III are set to be $\alpha=2000$, $L=1500$, $M=8$, $\xi=10^{-7}$, where $\alpha$ and $\xi$ are set as those in \cite{dragomiretskiy2013variational, 8890883}. Subfigure (a) and (b) depict the power spectral density (PSD) of each tone, where different tones are depicted in different colors, and the gray dashed line represents the PSD of the mixed signal in the noiseless case. 
It is obvious in subfigure (a) that at 30dB, both VMD and JVMD are able to precisely separate the three signal components with distinct tones. However, at 10dB, the center frequencies of the signal components separated by VMD deviate from the actual frequencies. In contrast, JVMD still accurately separates each signal component.
Besides, we provide a more intuitive comparison of the average center frequency estimation errors of VMD and JVMD in subfigure (c) when SNR varies, where each point is obtained by averaging over 1000 trials.
Apparently, the estimation error of JVMD enjoys a reduction of about ten times relative to VMD, indicating its robustness to noise.

\begin{figure}[htbp]
\centerline{\includegraphics[width=\columnwidth]{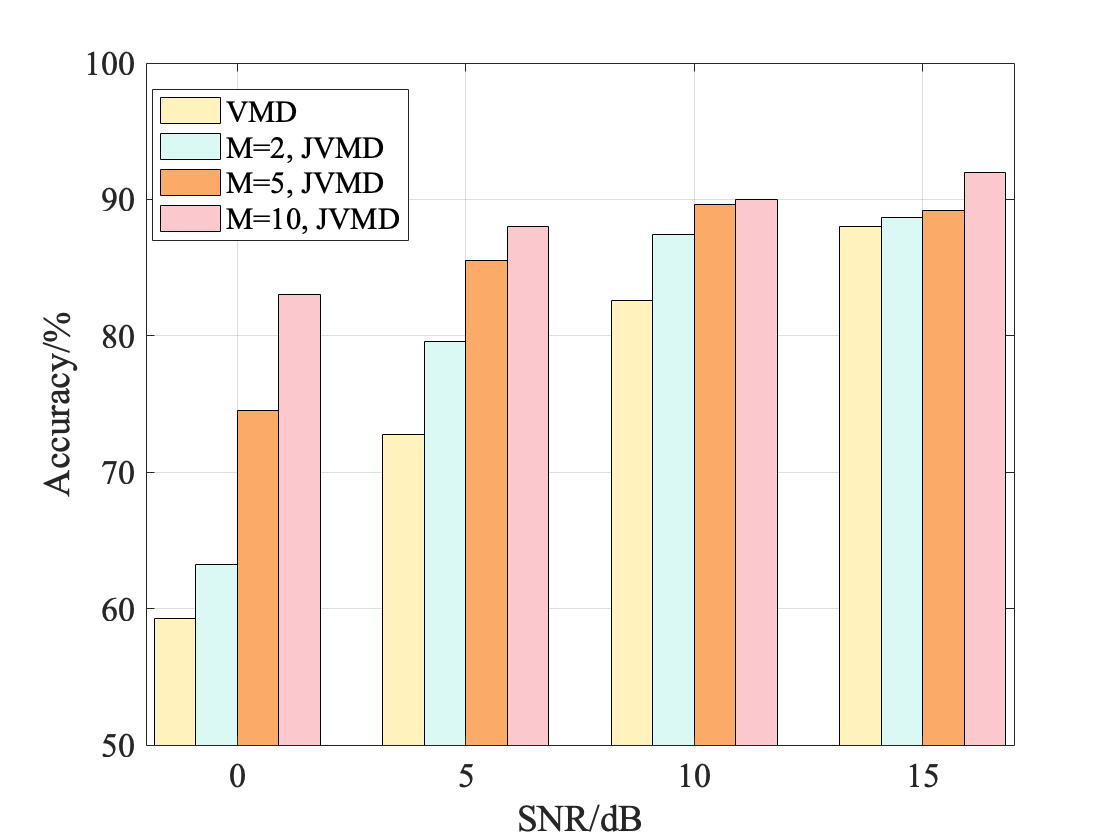}}
\captionsetup{font={footnotesize}}
\caption{VMD and JVMD-based SEI results using SRC classification with different SNRs.}
\label{sei1}
\end{figure}
\begin{figure}[htbp]
\centerline{\includegraphics[width=\columnwidth]{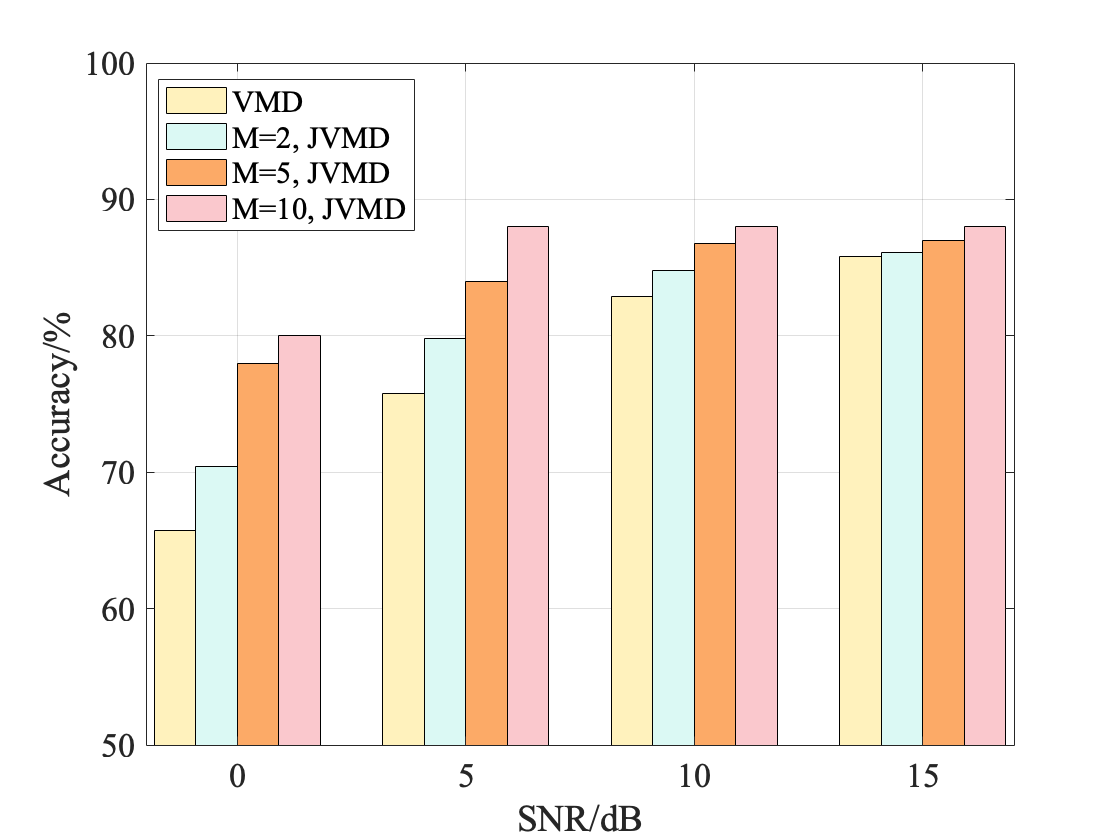}}
\captionsetup{font={footnotesize}}
\caption{VMD and JVMD-based SEI results using LC-KSVD classification with different SNRs.}
\label{sei2}
\end{figure}
\subsection{SEI based on JVMD}
In this subsection, we use the ADS-B dataset, which is a large-scale real-world radio signal dataset provided by \cite{ya2022large}, to evaluate our proposed JVMD-based SEI scheme.
The ADS-B dataset consists of 10 categories with 1000 test samples and 3080 training samples, with each sample being 4800 in length.
We select 100 samples for each category as the training set.
Different from tone separation, in order to preserve as much signal detail as possible, $\alpha$ in both the VMD and JVMD is set to be 500 in this experiment.
Moreover, we manually added different levels of noise to this ADS-B dataset to examine the performance of our proposed scheme in channels with different qualities.

Fig. \ref{sei1} and Fig. \ref{sei2} show the recognition results of VMD- and JVMD-based SEI schemes when using SRC and LC-KSVD as classification methods, respectively. 
The number of joint frames in the JVMD algorithm is chosen to be 2, 5, or 10.
As shown in these figures, the performance of the scheme with JVMD outperforms the one with VMD in all cases, and its accuracy improvement is especially noticeable at low SNRs.
In particular, the more frames are joined, the more robust JVMD is to noise. It can be seen that when $M=10$, the classification accuracy of JVMD-based SEI is higher than 80\% for all SNRs ranging from 0 to 15dB.
\section{Conclusion}
This paper studies the VMD-based SEI, where JVMD is proposed to extract signal features for subsequent emitter classification. JVMD exploits the stability of device fingerprints among successive frames of signals and preserves as many signal details as possible while reducing the effect of noise during the modal decomposition.
Besides, JVMD exhibits reduced complexity in comparison with VMD. 
Simulations on JVMD and VMD show the improved performance of the former in terms of the estimated center frequency of each tone. 
Furthermore, the simulations of the overall SEI scheme with JVMD for real-world datasets illustrate that the joint operation of multi-frame signals brings significant improvement compared with VMD in device classification accuracy, especially at low SNRs. 
\section*{Acknowledgment}

This work was supported by the National Natural Foundation of China (62371053) and by the US National Science Foundation (2136202).


\bibliographystyle{IEEEtran}
\bibliography{Ref} 
\end{document}